\documentstyle[preprint,epsfig,aps]{revtex}

\draft
\begin{document}
%\wideabs{
\title{Coexistence of bond-order wave and antiferromagnetism in a 
two-dimensional half-filled Peierls-Hubbard model}
\author{Qingshan Yuan$^{1,2}$ and Thilo Kopp$^1$}
\address{$^1$ Experimentalphysik VI, Universit\"at Augsburg, 
 86135 Augsburg, Germany\\
$^2$Pohl Institute of Solid State Physics, Tongji University, 
Shanghai 200092, P.R.China}

\maketitle

\begin{abstract}
The two-dimensional Peierls-Hubbard model is studied at half-filling 
within both Hartree-Fork and Kotliar-Ruckenstein slave-boson theory. 
The interplay between two types of long-range order, bond-order wave (BOW) 
and antiferromagnetism (AFM), is analysed for two representative
dimerization patterns, corresponding both to the same 
wavevector $(\pi,\pi)$. For each pattern, 
the Peierls dimerization (and associated BOW) is weakened and finally 
suppressed with increasing Hubbard on-site interaction, 
and correspondingly AFM is gradually enhanced. 
In particular, a coexistence regime with both BOW and AFM order 
is established in the parameter space of electron-lattice and 
Hubbard interactions. 
\end{abstract}


%}

\section{Introduction}

The Peierls instability towards spatially broken symmetry is 
an important phenomenon in low dimensional materials \cite{Peierls}. 
The one-dimensional (1D) case has been widely discussed in the context of 
polyacetylene (CH)$_x$ based on the Su-Schrieffer-Heeger (SSH) model 
\cite{SSH,Gruner}, where lattice displacements couple to electron 
hopping. For a half-filled band arbitrary small electron-lattice (e-l) 
coupling will induce a lattice dimerization (disregarding quantum lattice 
fluctuations), which is associated with a periodic
modulation of the bond hopping, that is, a so called on-bond charge-density 
wave or bond-order wave (BOW) \cite{Supp}. It has been established
that the Hubbard on-site Coulomb electron-electron (e-e) interaction $U$
will enhance the bond alternation initially for small values and finally 
suppress it at large values of $U$ \cite{Kivelson,Dixit,Vogl}.

\begin{figure}
\centerline{\epsfig{file=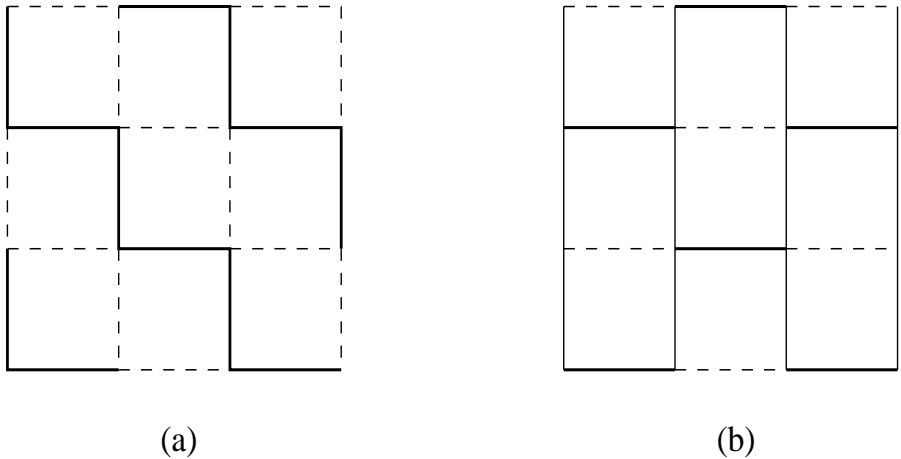,width=12cm,height=6cm}}
%\epsfxsize=12cm
%\epsfysize=6cm
%\centerline{\epsffile{Pattern.eps}}
\medskip
\caption{The lattice distortion patterns (a) and (b). In the figure 
a thick solid line 
corresponds to a strong bond with hopping integral $t(1+\delta)$, a dashed 
line corresponds to a weak bond with hopping integral $t(1-\delta)$, and a 
thin solid line corresponds to a normal bond with hopping integral $t$. Both 
patterns correspond to phonons with wave vector $(\pi,\pi)$. The dimerization 
is along two axes for pattern (a), while only along the $x$ axis 
for pattern (b).}
\label{Fig:Pattern}
\end{figure}

In two dimensions few theoretical investigations exist 
\cite{Vogl,Tang,Mazumdar,Clay,Yuan}, some of which connect the physics of  
Peierls systems to that of the high-$T_c$ copper 
oxides \cite{Tang,Mazumdar}. Moreover, these investigations may be of direct 
relevance to those quasi-two dimensional (2D) materials which 
show a Peierls instability such as 
transition-metal oxide bronzes like AMo$_6$O$_{17}$ (A=Na, K, Tl) 
\cite{Schlenk} and organic conductors like (BEDT-TTF)$_2M$Hg(SCN)$_4$ 
($M$=K, Rb, Tl) \cite{Sasaki,Clay}. As an effective minimal model, 
in this context, the 2D version of the SSH model was investigated 
\cite{Tang,Mazumdar,Yuan}. With only nearest-neighbour (n.n.) hopping on 
a square lattice, the electronic Fermi surface is perfectly nested at 
half-filling with nesting vector $Q=(\pi,\pi)$. Two possible alternation 
patterns for the lattice distortion and the concurrent bond hopping 
comply with this $Q$, as illustrated in Fig.~\ref{Fig:Pattern}. 
Whereas for pattern (a) the dimerization is in both
directions, it is only in one direction for pattern (b) \cite{Tang}.
Similar to the 1D case, already for an
infinitesimal e-l coupling, the 2D SSH model goes through 
a Peierls instability into one of the dimerized states 
of Fig.~\ref{Fig:Pattern} \cite{Yuan}. 

When a Hubbard on-site Coulomb interaction $U$ is included --- 
the model is then the so called Peierls-Hubbard model, results differ from 
the 1D case. Numerical calculations on a small 2D lattice \cite{Tang,Mazumdar} 
indicated that the Peierls instability
will be suppressed by the Hubbard $U$ \cite{Liu}. 
An intuitive explanation is that the on-site Coulomb interaction favors 
a spin-density wave (SDW) long-range order, that is, antiferromagnetism (AFM), 
while the dimerization associated with BOW harmonizes with a spin-singlet 
formation between those two n.n. spins which are connected by a strong bond.
As we know, due to the same nesting effect, the pure 2D half-filled Hubbard 
model (without consideration of a Peierls instability) 
has been shown to exhibit AFM long-range order for any $U>0$. 
This is in stark contrast to the corresponding 1D case where no true 
long-range AFM order is available and the correlated state rather corresponds
to a resonant valence bond state with strong weight 
from n.n. singlets \cite{Anderson}.
Consequently, one may envisage, for finite e-l coupling 
(denoted as $\eta$, see below) and e-e on-site interaction $U$, 
a competition between BOW and AFM as the 
underlying physics in the 2D half-filled Peierls-Hubbard model.
In the large $U$ limit, Zhang and Prelovsek have studied 
the corresponding spin-Peierls (SP) instability and found that the SP state,
competing with AFM, does not appear unless the spin-lattice coupling 
(analogous to $\eta$ here) exceeds a threshold \cite{Zhang}. 

The details of the competition between the two ordered states were studied
only for the above limiting case and the situation is not clear
for general values of $\eta$ and $U$. In particular, a basic problem
has to be solved: does BOW disappear once the AFM order sets in, or is
a coexistence of the two orders possible? It was previously argued by 
Mazumdar within a real space approach that the appearance of the AFM should 
coincide with the disappearance of the BOW \cite{Mazumdar}, 
which was, however, not verified.
To clarify this issue --- which is the topic of this paper, 
one needs to explicitly calculate the 
two order parameters for the BOW and AFM with varying $\eta$ and/or $U$. 
%Obviously, the study of the interplay of e-l
%and e-e interactions in a 2D system is of general theoretical interest
%\cite{Fehske}, and moreover may be connected to the real materials 
%mentioned above.

In this paper we make use of both the Hartree-Fock (HF) and 
Kotliar-Ruckenstein slave-boson (SB) approach \cite{Kotliar} to treat the 
Hubbard interaction. The HF results are usually valid at weak-coupling, 
and they can be used as a basis for further elaborate 
studies. In the context of investigations on density wave instabilities, 
the HF theory may give reasonable results even in one 
dimension \cite{Fukuyama}, where one would expect it to be worst 
because of strong fluctuations.
In dimensions higher than one, as considered here, qualitatively correct
results are expected from the HF theory. 
In order to extend the controlled weak-coupling results to intermediate
values of $U$, we evaluate BOW and AFM within a slave boson mean-field approach,
which is considered to be appropriate to interpolate from weak to 
strong coupling \cite{Fresard}.

The paper is organized as follows. In the next section the model Hamiltonian 
is introduced, and then the HF and SB approaches are formulated.
The self-consistent equations for the order parameters are derived in both
theories. In Sec. III numerical results are presented. The main results are 
shown in Fig.~\ref{Fig:OP}, where the coexistence of BOW and AFM
is found to be possible for each of the two patterns, in contrast to Mazumdar's 
argument. A complete comparison is made between the results derived 
from both approaches. Conclusive remarks are 
given in Sec. IV. An appendix completes the SB formulation.

\section{Formulation}
We begin with the 2D half-filled Peierls-Hubbard model
\begin{eqnarray}
H & = & H_t+H_U+H_K \label{eq:H}
\end{eqnarray}
with
\begin{eqnarray*}
H_t & = & -t\sum_{i,j,\sigma} [1+\alpha (u_{i,j}^x-u_{i+1,j}^x)]
  (c_{i,j,\sigma}^{\dagger}c_{i+1,j,\sigma}+{\rm h.c.})\\
& & -t\sum_{i,j,\sigma} [1+\alpha (u_{i,j}^y-u_{i,j+1}^y)]
  (c_{i,j,\sigma}^{\dagger}c_{i,j+1,\sigma}+{\rm h.c.}) \ ,\\
H_U & = & U\sum_{i,j} n_{i,j,\uparrow}n_{i,j,\downarrow} \ ,\\
H_K & = & {K\over 2} \sum_{i,j}[(u_{i,j}^x-u_{i+1,j}^x)^2+
(u_{i,j}^y-u_{i,j+1}^y)^2] \ ,
\end{eqnarray*}
where $c_{i,j,\sigma}^{\dagger} (c_{i,j,\sigma})$ is the creation 
(annihilation) operator for an electron at site $(i,j)$ with spin $\sigma$
($i$ denotes $x$ coordinate and $j$ denotes $y$ coordinate), $n_{i,j,\sigma}$
is defined as $n_{i,j,\sigma}=c_{i,j,\sigma}^{\dagger}c_{i,j,\sigma}$,
$u_{i,j}^{x/y}$ is the displacement component of site $(i,j)$ 
in $x/y$ direction, $t$ is the n.n. hopping parameter, and $\alpha$ is the 
electron-lattice coupling constant. $H_U$ is the Hubbard on-site interaction
with the repulsion strength $U$. The last term $H_K$ is the lattice 
elastic potential energy, with $K$ 
the elastic constant. The phonons are treated in adiabatic approximation.

For an analytical treatment on an infinite lattice, we have to work with a
definite distorted lattice, rather than allowing the distortions to arise 
arbitrarily. In this paper we constrain the discussion to the lattice 
distortions within the two commonly used dimerization patterns shown 
in Fig.~\ref{Fig:Pattern}. Only these patterns correspond to the nesting
vector $Q=(\pi,\pi)$ and they realize an unconditional Peierls instability 
that occurs for $\alpha \rightarrow 0$ and $U=0$. Explicitly they are written 
as  
$$u_{i,j}^x-u_{i+1,j}^x=(-1)^{i+j}u,\ \
  u_{i,j}^y-u_{i,j+1}^y=(-1)^{i+j}u
$$
for pattern (a) and
$$u_{i,j}^x-u_{i+1,j}^x=(-1)^{i+j}u,\ \  
  u_{i,j}^y-u_{i,j+1}^y=0\ \ \ \ \ 
$$
for pattern (b). For convenience, two dimensionless parameters
are defined: the dimerization 
amplitude $\delta=\alpha u$ and the electron-lattice coupling 
constant $\eta= \alpha^2 t/K$. Throughout the paper the hopping integral
$t$ is taken as the energy unit.

In the following we will construct the analytical formulas based on the HF 
and SB approaches, respectively, and leave the numerical calculations
to the next section.

\subsection{Hartree-Fock theory}
In our model, the on-site charge density wave is not favored and the total 
electron number on each site is uniform and equal to one at half-filling. 
Then the expectation 
value of the electron density with a given spin may be simply assumed
as $\langle n_{i,j,\sigma}\rangle ={1\over 2}[1+\sigma(-1)^{i+j}m]$ when 
the AFM order is taken into account, where $m$ represents the staggered 
magnetization. In HF approximation (equivalent to Hartree here) the 
local Hubbard term may be decoupled as 
$U\sum_{i,j} n_{i,j,\uparrow}n_{i,j,\downarrow}\rightarrow 
U\sum_{i,j}(n_{i,j,\uparrow}\langle n_{i,j,\downarrow}\rangle +
\langle n_{i,j,\uparrow}\rangle n_{i,j,\downarrow}-\langle n_{i,j,\uparrow}\rangle 
\langle n_{i,j,\downarrow}\rangle )$.
Then the Hamiltonian becomes quadratic and may be easily diagonalized in
momentum space. Under consideration of a bipartite lattice 
the final electronic spectra are derived as follows for pattern (a)
and (b), respectively,
\begin{eqnarray}
\varepsilon_{{\bf k},a}^{\pm} & = & \pm 
\sqrt{U^2m^2/4+4[(\cos k_x+\cos k_y)^2+\delta^2(\sin k_x+\sin k_y)^2]}\ ,
\label{eq:HFa} \\
\varepsilon_{{\bf k},b}^{\pm} & = & \pm 
\sqrt{U^2m^2/4+4[(\cos k_x+\cos k_y)^2+\delta^2\sin ^2 k_x]}\ . 
\label{eq:HFb}
\end{eqnarray}
Each branch above ($-$ or $+$) is two-fold (spin) degenerate.
The wave vector ${\bf k}=(k_x,k_y)$ is restricted to the reduced
Brillouin zone: $-\pi < k_x\pm k_y \le \pi$. With inclusion of constant terms
the ground state energy is
$$
E_\nu=2\sum_{\bf k} \varepsilon_{{\bf k},\nu}^- +NU(1+m^2)/4+E_{L,\nu}\ ,
$$
where $\nu=a,\ b$ represent pattern (a) and (b), respectively,
$N$ is the total number of lattice sites and
$E_{L,\nu}$ denote the
lattice elastic energies for both patterns: $E_{L,a}=2E_{L,b}=N\delta^2/\eta$.

The self-consistent equations for dimerization $\delta$ and 
magnetization $m$ are found by minimization of the ground state energy. 
They read: $\partial E_\nu/\partial \delta=0$ and 
$\partial E_\nu/\partial m=0$. 
The latter results in (except for a trivial solution $m=0$):
\begin{eqnarray}
1 & = & {U\over N}\sum_{\bf k} {1\over |\varepsilon_{{\bf k},\nu}^-|}\ ,
\label{eq:m}
\end{eqnarray}
and the former leads to
\begin{eqnarray}
1 & = & {4\eta \over N}\sum_{\bf k} {(\sin k_x+\sin k_y)^2\over 
|\varepsilon_{{\bf k},a}^-|}
\end{eqnarray}
for pattern (a) and 
\begin{eqnarray}
1 & = & {8\eta \over N}\sum_{\bf k} {\sin ^2 k_x \over 
|\varepsilon_{{\bf k},b}^-|}
\end{eqnarray}
for pattern (b).

\subsection{Slave-Boson theory}
In the spirit of the Kotliar-Ruckenstein SB approach \cite{Kotliar}, 
four auxiliary bosons
$e_{i,j}^{(\dagger)},\ p_{i,j,\sigma}^{(\dagger)} (\sigma=\uparrow,\downarrow),
\ d_{i,j}^{(\dagger)}$ are introduced to label the four different states for an
arbitrary site $(i,j)$, which can be empty, singly occupied by an electron
with spin up or down, or doubly occupied. The unphysical states in the
enlarged Hilbert space are eliminated by imposing two sets of local 
constraints:
\begin{eqnarray}
e_{i,j}^{\dagger} e_{i,j}+\sum_{\sigma} p_{i,j,\sigma}^{\dagger} 
p_{i,j,\sigma} +
d_{i,j}^{\dagger} d_{i,j} & = & 1 \ ({\rm completeness}) \ ,\label{eq:cons1}
\end{eqnarray}
and 
\begin{eqnarray}
p_{i,j,\sigma}^{\dagger} p_{i,j,\sigma} + d_{i,j}^{\dagger} d_{i,j} 
& = & c_{i,j,\sigma}^{\dagger} c_{i,j,\sigma}
\ ({\rm correctness\ of\ fermion\ occupancy\ for\ a\ given\ spin}) \ .
\label{eq:cons2}
\end{eqnarray}
For a bipartite lattice, we introduce a set of bosons, with
separate Lagrange multipliers for each sublattice. At the mean-field
level, the bosons are replaced by c-numbers and assumed to be 
site-independent on each sublattice. At the same time, the constraints above
are softened to be satisfied only on the average on each sublattice. 
This treatment is equivalent to making a saddle-point approximation 
in the path-integral formulation. For concreteness, 
we introduce the following parametrization
for sublattice A (and similar parameters are defined for sublattice B):
$e_{A},\ p_{A\sigma},\ d_{A}$ as average values of the boson operators
$e_{i,j}^{(\dagger)},\ p_{i,j,\sigma}^{(\dagger)},\ d_{i,j}^{(\dagger)}$, 
respectively, and
$\lambda_{A},\ \lambda_{A}^{\sigma}$ as Lagrange multipliers associated 
with the constraints (\ref{eq:cons1}), (\ref{eq:cons2}), respectively.
Then the Hamiltonian (\ref{eq:H}) may be recast into the following form
[we choose pattern (a) as an example]:
\begin{eqnarray}
H & = & -t(1+\delta)\sum_{R_{i,j}\in A,\sigma} z_{\sigma}(a_{i,j,\sigma}
^{\dagger}b_{i+1,j,\sigma}+a_{i,j,\sigma}^{\dagger}b_{i,j+1,\sigma}+{\rm h.c.})
\nonumber\\
& & -t(1-\delta)\sum_{R_{i,j}\in B,\sigma} z_{\sigma}(b_{i,j,\sigma}
^{\dagger}a_{i+1,j,\sigma}+b_{i,j,\sigma}^{\dagger}a_{i,j+1,\sigma}+{\rm h.c.})
+NU(d_A^2+d_B^2)/2+E_{L,a}
\nonumber\\
& & -\lambda_A\sum_{R_{i,j}\in A}(e_A^2+
\sum_{\sigma}p_{A\sigma}^2+d_A^2-1)-\sum_{R_{i,j}\in A,\sigma}
\lambda_A^{\sigma}(p_{A\sigma}^2+d_A^2-a_{i,j,\sigma}^{\dagger}a_{i,j,\sigma})
\nonumber\\
& & -\lambda_B\sum_{R_{i,j}\in B}(e_B^2+
\sum_{\sigma}p_{B\sigma}^2+d_B^2-1)-\sum_{R_{i,j}\in B,\sigma}
\lambda_B^{\sigma}(p_{B\sigma}^2+d_B^2-b_{i,j,\sigma}^{\dagger}b_{i,j,\sigma})
\ , \label{eq:H_SB}
\end{eqnarray}
where $a^{\dagger}(a)$ and $b^{\dagger}(b)$ are the electron creation 
(annihilation) operators for sublattice A and B, respectively. The hopping 
renormalization factor $z_{\sigma}$ ensures the correct result in the limit 
of vanishing $U$ and takes the form $z_{\sigma}=\langle z_{\sigma}^A 
\rangle \langle z_{\sigma}^B\rangle$ with
$$
\langle z_{\sigma}^L\rangle={e_Lp_{L\sigma}+p_{L\bar{\sigma}}d_L \over 
\sqrt{(1-e_L^2-p_{L\bar{\sigma}}^2)(1-d_L^2-p_{L\sigma}^2)}}
\ \ \ \ \ \ \ L=A,\ B\ .
$$
The Hamiltonian (\ref{eq:H_SB}) may be diagonalized in momentum space and
the energy bands read:
\begin{eqnarray}
\epsilon_{{\bf k}\sigma,a}^{\pm} & = & (\lambda_A^{\sigma}+
\lambda_B^{\sigma})/2 \pm
\sqrt{(\lambda_A^{\sigma}-\lambda_B^{\sigma})^2/4+
4z_{\sigma}^2[(\cos k_x+\cos k_y)^2+\delta^2(\sin k_x+\sin k_y)^2]}\ .
\end{eqnarray}
Similarly, the energy bands for pattern (b) are given by:
\begin{eqnarray}
\epsilon_{{\bf k}\sigma,b}^{\pm} & = & (\lambda_A^{\sigma}+
\lambda_B^{\sigma})/2 \pm
\sqrt{(\lambda_A^{\sigma}-\lambda_B^{\sigma})^2/4+
4z_{\sigma}^2[(\cos k_x+\cos k_y)^2+\delta^2\sin^2 k_x]}\ .
\end{eqnarray}
At half-filling and zero temperature only the two lower ($-$) bands
are occupied (the constant $\lambda_A^{\sigma}+\lambda_B^{\sigma}$ is 
independent of $\sigma$ as will be seen later). Then the ground state energy
is expressed as ($\nu=a,\ b$)
\begin{eqnarray}
E_{\nu} & = & \sum_{{\bf k}\sigma}\epsilon_{{\bf k}\sigma,\nu}^- +E_0+ 
   E_{L,\nu}\ 
\end{eqnarray}
with the constant
$E_0=(N/2)[U(d_A^2+d_B^2)-\lambda_A (e_A^2+\sum_{\sigma}p_{A\sigma}^2+
d_A^2-1)-\lambda_B (e_B^2+\sum_{\sigma}p_{B\sigma}^2+d_B^2-1)-
\sum_{\sigma}\lambda_A^{\sigma}(p_{A\sigma}^2+d_A^2)-
\sum_{\sigma}\lambda_B^{\sigma}(p_{B\sigma}^2+d_B^2)]\ .
$

The self-consistent equations are obtained from the requirement that 
the ground state energy is stationary with respect to the parameters: 
$e_{A(B)},\ p_{A(B)\sigma},\ d_{A(B)},\ \lambda_{A(B)},\ \lambda_{A(B)}^
{\sigma},\ \delta$. Except for the equation corresponding to $\delta$, they 
all have the general form: $\sum_{{\bf k}\sigma}\partial 
\epsilon_{{\bf k}\sigma,\nu}^- /\partial X + \partial E_0/ \partial X=0$, where
$X$ represents one of the parameters. Analyzing these equations
and applying the constraint $\sum_{\sigma}p_{A(B)\sigma}^2+2d_{A(B)}^2=
\sum_{\sigma}\langle a(b)_{i,j,\sigma}^{\dagger}a(b)_{i,j,\sigma}\rangle=1$ 
at half-filling, one may find the solution
satisfying the following relations:
$e_A=e_B=d_A=d_B(=d)$, $\lambda_A=\lambda_B(=\lambda)$, 
$p_{A\sigma}=p_{B\bar{\sigma}}$,
$\lambda_A^{\sigma}=\lambda_B^{\bar{\sigma}}$ and $\sum_{\sigma}
\lambda_A^{\sigma}=\sum_{\sigma}\lambda_B^{\sigma}=U$. Consequently,
the number of free parameters is substantially reduced. 
The final compact self-consistent equations 
are listed in the Appendix with several re-defined independent 
parameters: $d$, $\lambda$, $\lambda_{AB}=\lambda_A^{\uparrow}-\lambda_B^{\uparrow}=\lambda_B^{\downarrow}-\lambda_A^{\downarrow}$, 
$m=p_{A\uparrow}^2-p_{A\downarrow}^2=p_{B\downarrow}^2-p_{B\uparrow}^2$,
where $m$ denotes the same staggered magnetization as in HF theory.

\section{Numerical Results and Discussions}
We now focus on the numerical results obtained from the
self-consistent equations. First, it is necessary to analytically analyze 
the mean-field equations more thoroughly.
After a replacement of the momentum summation by integration, i.e., 
$\sum_{\bf k} \rightarrow
{N\over 2\pi ^2}\int \int_{-\pi/2}^{\pi/2}\ {\rm d}k_1{\rm d}k_2$ 
$(k_{x,y}=\pm k_1+k_2)$, we examine Eq.~(\ref{eq:m}) in HF theory.
It is easily seen that the right-hand side (rhs) of Eq.~(\ref{eq:m}) assumes
a different analytical behavior for each of the two patterns due to their
different respective spectra. For pattern (a) the rhs is divergent at $m=0$, 
irrespective of the value of $\delta$,
and decreases monotonically with increasing $m$ until it reaches a value less 
than $1$ at $m=1$. This implies for pattern (a) that Eq.~(\ref{eq:m}) can 
be always solved with a nonzero solution of $m$ as long as $U>0$. 
For pattern (b) it is not 
always the case because the rhs is finite at $m=0$ for any $\delta >0$. 
Once this finite value is less than $1$, 
%Eq.~(\ref{eq:m}) has no solution and simply 
one has to adopt the trivial $m=0$ solution. 
The same conclusion may be obtained by a similar analysis of the 
corresponding equations in the SB evaluation [see Eqs. (\ref{eq:lamAB})
and (\ref{eq:SBm}) in the Appendix].

\begin{figure}[ht]
\centerline{\epsfig{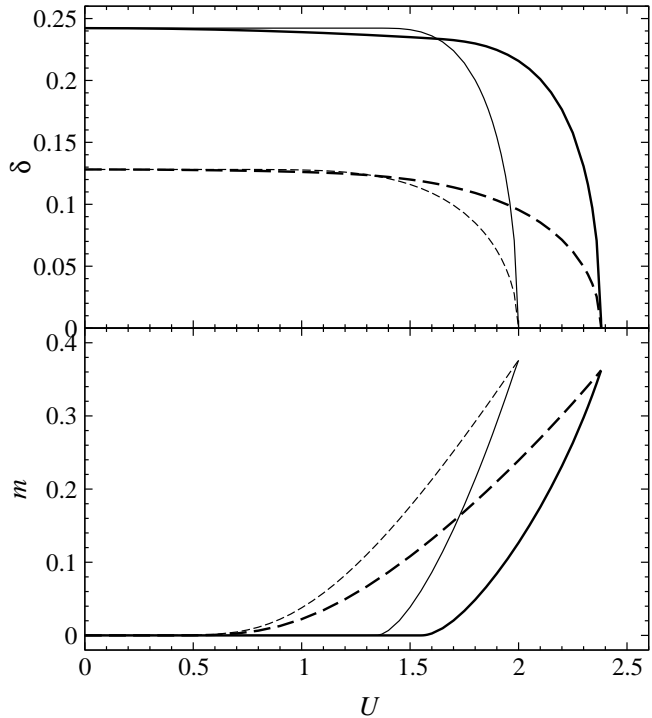}}
\smallskip
\caption{The optimal values for the dimerization $\delta$ and AFM $m$ 
as functions of $U$ at $\eta=0.5$. In each panel, 
the dashed lines are for pattern (a) and the 
solid ones are for pattern (b); the thick lines are the SB results and
the thin ones are the HF results.}
\label{Fig:OP}
\end{figure}

The numerical solutions for $\delta$ and $m$ are displayed as functions of $U$
in Fig.~\ref{Fig:OP} for both patterns and in both approaches, whereby the 
e-l coupling $\eta$ is fixed at $0.5$. 
Globally, it is seen that for both patterns $\delta$ tends to 
decrease and $m$ to increase with growing $U$. 
This supports the notion that the on-site interaction tends to favor
the AFM order and to suppress the Peierls dimerization. Let us go to the
details in the following.

In HF (see all the thin lines), it is found that for pattern (a) 
$m$ becomes finite (although small at small $U$)
and simultaneously $\delta$ begins to decrease once $U>0$; 
while for pattern (b) $m$ stays zero for small $U$ 
up to $U>U_m\simeq 1.34$ where it becomes finite and correspondingly, $\delta$ 
first keeps its $U=0$ value and then starts to decrease for $U>U_m$.
The dimerization disappears at the same critical value $U_c\simeq 2$ for 
both patterns and it approaches zero smoothly and quickly when $U$ is 
close to $U_c$. Comparing the results for $m$ vs. $U$ between the two patterns, 
it is clear that pattern (a) is more favorable to the formation of 
AFM order than pattern (b). We will come to this point later. 

Most of the above qualitative results are also found in SB approach
(see all the thick lines). On the other hand,
the difference to the HF results is clear as well, which we want to 
emphasize here. 
A distinct quantitative difference is that for each of the patterns
the AFM order derived from SB theory is (much) weaker than that from 
HF theory, and correspondingly the dimerization decreases to zero over a 
larger range of $U$. The critical value for the disappearance of $\delta$ 
now becomes $U_c\simeq 2.38$, and the
necessary Hubbard interaction to induce finite $m$ in the case of pattern (b)
is $U_m\simeq 1.56$. Both values are larger than the corresponding ones from
the HF theory, which is understandable. As is well known, the HF theory 
usually overestimates the tendency towards AFM order. The SB theory, as an 
improved approach to fluctuation contributions, should lead to a slower 
formation of AFM, which complies with our findings. 

A further important difference between the SB and HF results is observed 
in the region $U<U_m$ for pattern (b), 
where the AFM order has not yet formed. From the solid lines in the upper panel
of Fig.~\ref{Fig:OP}, it is seen that the dimerization $\delta$ keeps 
a constant value in HF theory, while it decreases slowly with increase of $U$ 
in SB theory. This disagreement may be understood as follows.
In HF theory, see e.g., Eq.~(\ref{eq:HFb}), the Hubbard $U$ becomes irrelevant
once the order parameter $m$ is zero: the value for $\delta$ will be the 
same as that without $U$. In the SB approach, however, the Hubbard $U$ is 
relevant even at $m=0$ by affecting the probability of double occupancy $d^2$. 
As $U$ increases, the double occupancy is disfavored, i.e., 
the quantity $d^2$ decreases. 
Correspondingly the effective hopping $tz_{\sigma}$ decreases too (cf. the 
expression for $W$ in the Appendix), which may be
understood equivalently as an increase of the elastic constant $K$ or
a reduction of the e-l coupling $\eta$. This signifies a decreasing dimerization.
In this point the HF theory fails to catch
the correct physics by assuming the probability of double occupancy as
a constant $1/4$ which is correct only for $U\rightarrow 0$.
\begin{figure}[ht]
\centerline{\epsfig{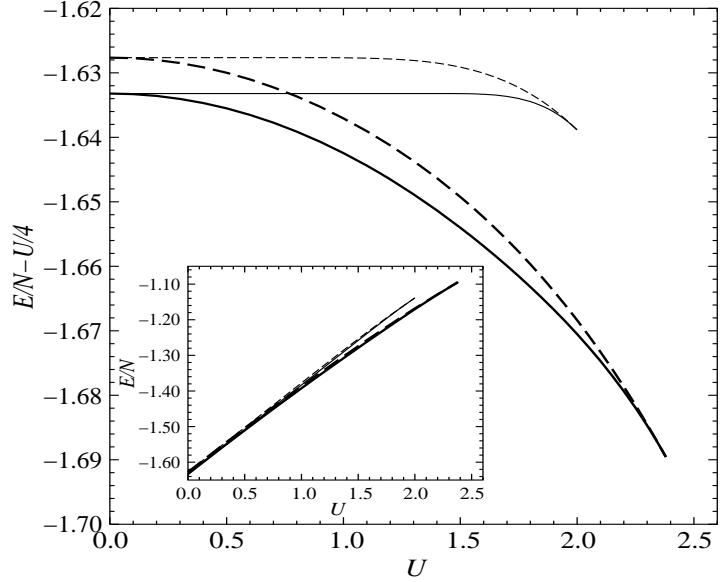}}
%\smallskip
\caption{The ground state energies (per site) as functions of $U$ at 
$\eta=0.5$, corresponding to Fig.~\ref{Fig:OP} (with the same line labels). 
Each line stops at its own critical point $U_c$. The original values of these 
energies are shown in the inset. All energies are in units of $t$.}
\label{Fig:Eg}
\end{figure}

Furthermore, we can check the stability between both patterns
by comparison of the ground state energies calculated in all cases, 
which are shown in Fig.~\ref{Fig:Eg}. 
For each pattern the SB approach gives a lower energy than the HF theory
in the whole range of $U$. Also, it is 
seen within each approach that pattern (b) has a lower energy than 
pattern (a), which signifies that pattern (b) is more stable. 

\begin{figure}[ht]
\centerline{\epsfig{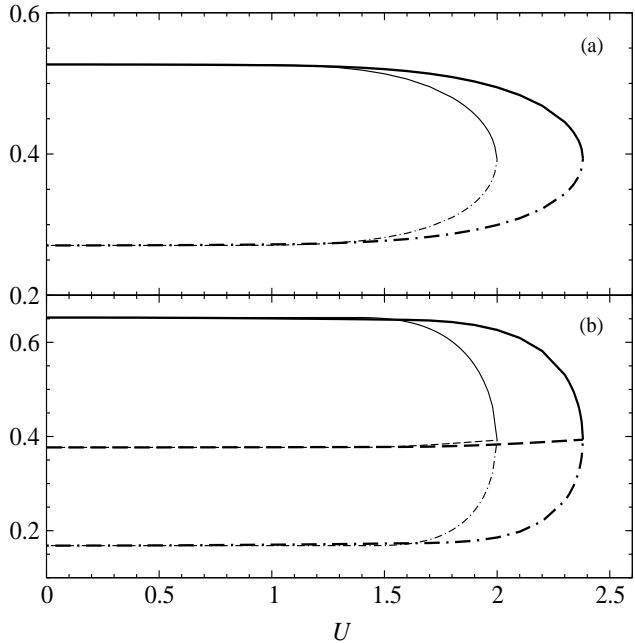}}
\smallskip
\caption{The expectation values $h_{1,2}^x$, $h_{1,2}^y$ as functions of $U$
for both patterns. 
In each panel, the solid lines show the quantity $h_1^x$
and the dot-dashed lines give $h_2^x$; the thick lines are the SB results and
the thin ones are the HF results. For pattern (a), $h_{1,2}^y=h_{1,2}^x$.
For pattern (b), $h_1^y=h_2^y$, which are plotted by the dashed lines 
in the lower panel.}
\label{Fig:BOW}
\end{figure}

It is worthwhile to point out that the BOW is always associated with 
the finite dimerization.
In order to see this, we have calculated the BOW which is characterized by a
modulation of the hopping amplitude. Explicitly, we define
the expectation values $h_1^x=\langle a_{i,j,\sigma}^{\dagger}b_{i+1,j,\sigma}+
{\rm h.c.}\rangle$, 
$h_2^x=\langle b_{i,j,\sigma}^{\dagger}a_{i+1,j,\sigma}+{\rm h.c.}
\rangle$ ($\sigma$ is irrelevant)
for the alternating bond hoppings along the $x$ axis, and similar values
$h_1^y$, $h_2^y$ for the hoppings along the $y$ axis.
By symmetry, we have $h_{1,2}^x=h_{1,2}^y$ for pattern (a) and 
$h_1^y=h_2^y$ for pattern (b). All the quantities are calculated in both 
theories and plotted as functions of $U$ in Fig.~\ref{Fig:BOW}. 
The BOW is exhibited by the inequality between $h_1^x$ and $h_2^x$ 
for each pattern \cite{supp}. It is clear for each approach that such an 
inequality is present within $0<U<U_c$, 
the same region where the dimerization is finite.

Numerically, Tang and Hirsch \cite{Tang} studied the same model and calculated 
the energy gain from dimerization as a function of $U$ for the pattern (b)
shown here. By studying how the energy gain changes with $U$, 
they found originally that the Hubbard $U$ has little effect on 
the dimerization until it is large enough to suppress it,
and later corrected that the dimerization is disfavored as soon as $U$ is 
present. The finite-size effect was cautioned 
by the authors themselves. Their principal result, i.e., the Hubbard $U$ is 
unfavorable to dimerization, is consistent with ours, especially with 
the SB results for pattern (b). 
Although it seems that the suppression of dimerization by $U$
is faster in our results than what they displayed, no direct comparison 
is available because they calculated neither the order parameters nor 
the ground state energies.
Obviously, further numerical calculations on large size systems 
are necessary for better comparison.

\begin{figure}[ht]
\epsfxsize=8cm
\epsfysize=10cm
\centerline{\epsfig{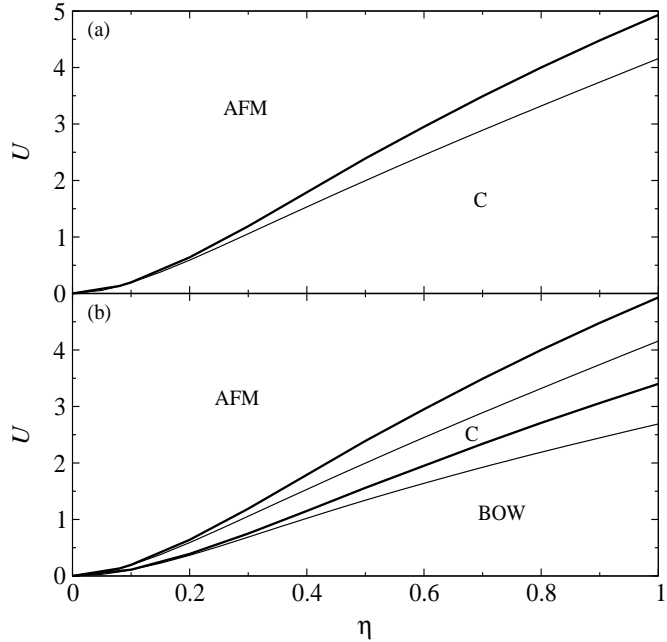}}
\smallskip
\caption{Stable phases for BOW, AFM, and the coexisting state (C) 
in the parameter space $(U,\eta)$ for pattern (a) [upper panel] and
pattern (b) [lower panel]. The thick
lines show the SB results and the thin ones show the HF results.}
\label{Fig:Ph}
\end{figure}

The main contribution in our work is that the order parameters as functions 
of $U$ are explicitly obtained so that the interplay between BOW and AFM 
becomes transparent. The problem proposed in the Introduction is then naturally
answered. It is clearly seen in Fig.~\ref{Fig:OP} that the BOW and AFM may 
coexist for both patterns. For pattern (a) the coexistence 
(i.e., $\delta>0,\ m>0$) 
appears as long as $U>0$ and pure AFM order exists for $U>U_c$. 
For pattern (b) the coexistence is limited to the region 
$U_m<U<U_c$. These results are not favorable to the argument by Mazumdar that 
the AFM emerges with the disappearance of the BOW \cite{Mazumdar}.
In fact, the valence bond approach adopted by Mazumdar in real space
is appealing. It states that, in order to implement a symmetry-broken state 
(e.g., BOW), ``extreme configurations'' with shortest
repeat units have to be identified. For 2D systems, he chose pattern (a) as 
the realization of the Peierls state and argued that 
the extreme configuration for BOW is a combination of zigzag chains and
the n.n. sites within each chain are doubly occupied and unoccupied, respectively.
However, the pattern selected in his work is not the 
pattern with the lowest energy and furthermore, the considered extreme 
configuration is actually disfavored, even for
weak Coulomb on-site interaction, as is verified in the exact $U\rightarrow 0$
approach of this paper. It implies that spin fluctuations are more
pronounced than charge fluctuations in the BOW state of the half-filled system.

Eventually, we determine the coexistence regions for different $\eta$
in both theories. The results are shown in Fig.~\ref{Fig:Ph}, 
where different phases are indicated 
in the parameter plane $(U,\eta)$. For pattern (a), only two phases exist, 
either a state with coexisting BOW and AFM or a pure AFM state. 
However for pattern (b) 
pure BOW and AFM states exist, which are separated by a coexisting state
--- the region between two thick (SB) or thin (HF) lines 
in Fig.~\ref{Fig:Ph}. As for the methods, globally speaking, 
the SB approach pushes the AFM order to the higher $U$ 
regime than the HF theory. 

%In the end of this paragraph,
%we mention that a similar coexistence between
%BOW and SDW was found for the same model but in one dimension within
%HF theory \cite{Kivelson,Xu}, although the 1D HF results incorrectly 
%implied long-range spin correlations.

Finally, we come back to the difference between the results for the two patterns. 
As explained above, pattern (a) is more favorable to the development of AFM
than pattern (b). It may be roughly understood from their different 
dimerization structures.
%As we know, the competition between the BOW and AFM depends on how the spin
%singlets 
As seen from Fig.~\ref{Fig:Pattern}, for pattern (b), 
each site is connected to only one n.n. site by a strong 
bond when the BOW (or dimerization) forms. Thus a spin singlet on this strong 
bond is apt to prevail in presence of $U$, which will resist the AFM.
On the other hand, for pattern (a), each site connects two n.n. sites 
with strong bonds. This, on the contrary, makes the construction of 
spin singlets on these strong bonds difficult and the AFM is 
easier to develop. 

\section{Conclusion}

We have investigated the Peierls-Hubbard model in two dimensions
at half-filling within both HF and SB approach. Two dimerization patterns,
corresponding to the same wavevector $(\pi,\pi)$, are considered and 
the interplay between two long-range order states, BOW and AFM, is addressed. 
For each pattern, it is found that the Peierls dimerization 
(and associated BOW) is weakened by the on-site interaction $U$
as soon as $U$ is present and finally suppressed at a critical $U=U_c$. 
Correspondingly, the AFM is favored by $U$. Whereas for pattern (a),
see Fig.~\ref{Fig:OP}, AFM is induced once $U>0$, it is
not activated until $U=U_m$ for pattern (b). For both patterns, the
coexistence of BOW and AFM is possible. 
SB and HF evaluations lead mostly to the same qualitative results and 
quantitatively the former approach results in larger values of $U_c$ and $U_m$.
Whereas the HF evaluation provides us with the exact weak coupling 
($U\rightarrow$ 0) result, the SB approach extends the findings to intermediate
coupling, and corrects charge and spin fluctuation contributions beyond HF. 
Especially, the reduction of charge fluctuations by $U$ decreases the 
dimerization $\delta$ consistently in the region $U<U_m$ for pattern (b).

\section*{Acknowledgements}

We would like to thank T. Nunner for valuable discussions.
This work was financially supported by the Deutsche Forschungsgemeinschaft
through SFB 484. Q. Yuan also acknowledges S. Mazumdar, H. Zheng for useful 
communications and the support by the National Natural Science Foundation of
China.

\appendix
\section*{Self-consistent equations in the SB theory}

In this appendix we implement the formulation of the SB theory. With respect to
the parameters $d,\ \lambda,\ \lambda_{AB},\ m,\ \delta$, the self-consistent
equations are derived as follows for pattern (a):
\begin{eqnarray}
d & = & -2C_1 I_2 / \lambda\ ,\label{eq:SBd}\\
-\lambda & = & U/2 + 2({C_2 / \sqrt{(1+m)/2-d^2}}+
{C_3 / \sqrt{(1-m)/2-d^2}})I_2
\ , \label{eq:lam} \\
\lambda_{AB} & = & 4({C_3 / \sqrt{(1-m)/2-d^2}}-
{C_2 / \sqrt{(1+m)/2-d^2}})I_2
\ ,\label{eq:lamAB}\\
m &= & -\lambda_{AB} I_1\ ,\label{eq:SBm}\\
1 & = & 4\eta W I_3 \ , \label{eq:SBdel}
\end{eqnarray}
where 
\begin{eqnarray*}
I_1 & = & {1\over N} \sum_{\bf k} {1 \over 
  \sqrt{\lambda_{AB}^2/4+4W[(\cos k_x+\cos k_y)^2+
  \delta^2(\sin k_x+\sin k_y)^2]}}\ ,\\
I_2 & = & {1\over N} \sum_{\bf k} {(\cos k_x+\cos k_y)^2+
   \delta^2(\sin k_x+\sin k_y)^2 \over \sqrt{\lambda_{AB}^2/4+
   4W[(\cos k_x+\cos k_y)^2+\delta^2(\sin k_x+\sin k_y)^2]}}\ ,\\
I_3 & = & {1\over N} \sum_{\bf k} {(\sin k_x+\sin k_y)^2 \over 
   \sqrt{\lambda_{AB}^2/4+4W[(\cos k_x+\cos k_y)^2+
   \delta^2(\sin k_x+\sin k_y)^2]}}\ ,\\
W & = & z_{\sigma}^2 = {16d^4(\sqrt{(1+m)/2-d^2}+\sqrt{(1-m)/2-d^2})^4
   \over (1-m^2)^2}\ ,\\
C_1 & = & {64d^3(\sqrt{(1+m)/2-d^2}+\sqrt{(1-m)/2-d^2})^3 \over (1-m^2)^2}\times \\
 & &  \ \ \ \left\{{\sqrt{(1-m)/2-d^2}[(1-m)/2+d^2]+d^2\sqrt{(1+m)/2-d^2}\over 1-m}+
    m\rightarrow -m \right\}\ ,\\
C_2 & = & {64d^4(\sqrt{(1+m)/2-d^2}+\sqrt{(1-m)/2-d^2})^3 (1-d^2+
   \sqrt{(1+m)/2-d^2}\sqrt{(1-m)/2-d^2}) \over (1-m^2)^2(1-m)}\ ,\\
C_3 & = & {64d^4(\sqrt{(1+m)/2-d^2}+\sqrt{(1-m)/2-d^2})^3 (1-d^2+
  \sqrt{(1+m)/2-d^2}\sqrt{(1-m)/2-d^2}) \over (1-m^2)^2(1+m)}\ .
\end{eqnarray*}
The equations for pattern (b) are the same as those above except that 
the expression $(\sin k_x+\sin k_y)^2$ in $I_{1,2,3}$
is substituted by $\sin ^2 k_x $, and the Eq.~(\ref{eq:SBdel}) 
is changed into
\begin{eqnarray}
1 & = & 8\eta W I_3\ .
\end{eqnarray}
Correspondingly, the ground state energies may be written in the simple form:
$$
E_a =-2\sum_{\bf k} \sqrt{\lambda_{AB}^2/4+4W[(\cos k_x+\cos k_y)^2+
\delta^2(\sin k_x+\sin k_y)^2]} +NUd^2-Nm\lambda_{AB}/2+E_{L,a}$$
for pattern (a) and
$$
E_b =-2\sum_{\bf k}\sqrt{\lambda_{AB}^2/4+4W[(\cos k_x+\cos k_y)^2+
\delta^2\sin ^2 k_x]} +NUd^2-Nm\lambda_{AB}/2+E_{L,b} 
$$
for pattern (b).

We point out that for both patterns there always exists a trivial 
solution with $m=0$ [consider Eqs. (\ref{eq:lamAB}) and (\ref{eq:SBm}) and 
note that $C_2=C_3$ at $m=0$].
Moreover, we checked that for $\delta=0$ the results presented by 
Fr\'esard {\it et al.} (e.g., staggered magnetization, ground state 
energy) for the pure Hubbard model \cite{Fresard} are reproduced.

%end{references}

\begin{thebibliography}{99}
\bibitem{Supp} For a Holstein-type (i.e., intrasite) electron-lattice 
coupling model, the lattice dimerization is associated with an 
on-site charge density wave.
\bibitem{supp} Actually, if one defines a BOW order parameter ${1\over N}
 \sum_{i,j} (-1)^{i+j} \langle c_{i,j,\sigma}^{\dagger}c_{i+1,j,\sigma}
 +{\rm h.c.}\rangle$, it is equal to $(h_1^x-h_2^x)/2$.
\end{thebibliography}
\end{document}